\documentclass[twocolumn,preprintnumbers,superscriptaddress,nofootinbib,aps,prd,floatfix]{revtex4}

\usepackage{graphicx}
\usepackage{amsmath,mathenv,bbm,amsfonts,color}
\usepackage{subfigure}
\usepackage{graphicx}
\usepackage{cases}
\usepackage{enumerate}
\usepackage{enumitem}
\usepackage{slashed}

\newenvironment{cedescription}{%
   
   \begin{description}[leftmargin=0.5cm, style=sameline]%
}{%
   \end{description}%
}

\newcommand{\Tr}{\text{Tr}}
\newcommand{\tr}{\text{Tr}}
\newcommand{\gev}{\text{GeV}}
\newcommand{\Br}{\text{Br}}

\newcommand{\gl}[1]{\eqref{#1}}

\newcommand{\alfa}{\alpha}

\newcommand{\dpm}{{\pm\pm}}
\newcommand{\dmp}{{\mp\mp}}

\newcommand{\SU}{\text{SU}}

\newcommand{\U}{\text{U}}

\newcommand{\nn}{\nonumber}

\begin{document}

\title{Pinning down Higgs triplets at the LHC}

\author{Christoph Englert} \email{christoph.englert@glasgow.ac.uk}
\affiliation{Institute for Particle Physics Phenomenology, Department
  of Physics,\\Durham University, Durham DH1 3LE, United Kingdom}
\affiliation{SUPA, School of Physics and Astronomy, University of
  Glasgow,\\Glasgow, G12 8QQ, United Kingdom}
\author{Emanuele Re} \email{emanuele.re@physics.ox.ac.uk}
\affiliation{Rudolf Peierls Centre for Theoretical Physics, Department
  of Physics,\\University of Oxford, Oxford, OX1 3NP, United Kingdom}
\author{Michael Spannowsky} \email{michael.spannowsky@durham.ac.uk}
\affiliation{Institute for Particle Physics Phenomenology, Department
  of Physics,\\Durham University, Durham DH1 3LE, United Kingdom}

\begin{abstract}
  Extensions of the Standard Model Higgs sector involving weak
  isotriplet scalars are not only benchmark candidates to reconcile
  observed anomalies of the recently discovered Higgs-like particle,
  but also exhibit a vast parameter space, for which the lightest
  Higgs' phenomenology turns out to be very similar to the Standard
  Model one. A generic prediction of this model class is the
  appearance of exotic doubly charged scalar particles. In this paper
  we adapt existing dilepton+missing energy+jets measurements in the
  context of SUSY searches to the dominant decay mode $H^{\pm\pm}\to
  W^\pm W^\pm$ and find that the LHC already starts probing the
  model's parameter space.  A simple modification towards signatures
  typical of weak boson fusion searches allows us to formulate even
  tighter constraints with the 7 TeV LHC data set. A corresponding
  analysis of this channel performed at 14 TeV center of mass energy
  will constrain the model over the entire parameter space and
  facilitate potential $H^\dpm \to W^\pm W^\pm$ discoveries.
\end{abstract}

\preprint{DHCP/13/90}
\preprint{IPPP/13/45}
\preprint{OUTP-13-12P}

\maketitle 


\section{Introduction}
\label{sec:intro}
The recent discovery~\cite{:2012gk,:2012gu,newboundsa,newboundsb} of
the Higgs boson~\cite{orig} provides an opportunity to check the
phenomenological consistency of various scenarios of electroweak
symmetry breaking with measurements for the first time. Higgs triplet
models have received considerable attention recently as they can
reconcile the possibly observed anomaly in the $H\to \gamma\gamma$
channel~\cite{lit,andrew,us,Killick:2013mya}. Whether this excess
persists or future measurements of the diphoton partial decay width
will return to the Standard Model (SM) values as suggested by recent
CMS results \cite{cmsaa} is unclear at the moment. However, as
demonstrated in~\cite{us}, there are certain models with Higgs
triplets~\cite{Georgi:1985nv,chano} which posses a large parameter
space where the resulting phenomenology is SM-like
\cite{carmi,Belanger:2013xza} even for larger triplet vacuum
expectation values. A generic prediction of electroweak precision
measurements in this case is the appearance of doubly charged scalar
particles $H^\dpm$ with a mass of several hundred GeV that result from
the weak triplet structure in the Higgs sector extension.

Due to the quantum numbers of the $\SU(2)_L$ triplet, Majorana
mass-type operators can induce a prompt decay of $H^\dpm$ into two
leptons with identical charge \cite{classic}. This interaction has
already been constrained at the LHC in multilepton
searches~\cite{searchhpp}.  However, as soon as the mass of the doubly
charged scalar exceeds twice the $W$ mass, the decay to gauge bosons
is preferred. This can be seen from the scaling of the partial decay
widths: $\Gamma(H^{\pm\pm}\to W^\pm W^\pm)/\Gamma(H^{\pm\pm}\to
\ell^\pm \ell^\pm) \sim m_{H^{\pm\pm}}^2/m_W^2$. Over the bulk of the
parameter space this leads to a dominant decay of the doubly charged
Higgs to $W$ bosons~\cite{Gunion:1989ci}. Formulating a meaningful
constraint of this model class must therefore not neglect $H^\dpm\to
W^\pm W^\pm$~\cite{Kanemura:2013vxa}.

\begin{figure}[!b]
  \centering
  \includegraphics[width=0.23\textwidth]{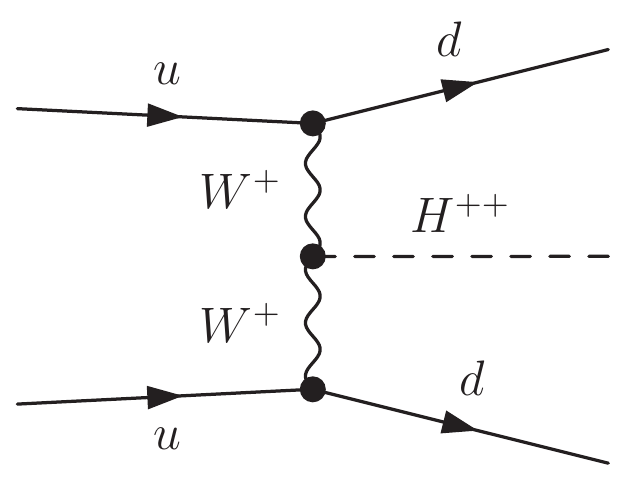}
  \caption{\label{fig:graph} Sample weak boson fusion diagram involved
    in the production of $H^\dpm$. We do not show the $H^\dpm$
    decay. By crossing one of the up-flavor quarks to the final state
    and the non-connected down-flavor to the initial we recover the
    Drell-Yan-type production modes.}
\end{figure}

The production of single intermediate $H^\dpm$ boson can only proceed
via weak boson fusion (WBF) diagrams (Fig.~\ref{fig:graph}) and
crossed processes ({\it{i.e.}} Drell-Yan type production). Hence,
$H^\dpm$ production inherits all the phenomenological advantages of
WBF Higgs and diboson production~\cite{dieter}. Producing the
relatively heavy final state requires energetic initial state
partons. The $t$-channel color singlet exchange results in relatively
small scattering angles of the two outgoing jets at moderate
transverse momentum and a central detector region essentially free of
QCD radiation. Eventually, the typical signature is two isolated
central leptons and missing energy, and two forward jets at large
rapidity differences with high invariant mass. Phenomenological
investigations of this signatures are helped by small irreducible SM
backgrounds~\cite{barbara,giulia}.  These signatures have already been
investigated partially in Refs.~\cite{vbftriplet,Chiang}, however
neither including a parameter scan involving the Higgs candidate's
signal strengths nor constraints from electroweak precision data
(EWPD).

To our knowledge, neither ATLAS nor CMS have performed a dedicated
analysis of this final state in the triplet Higgs model
context. However, there are searches for Supersymmetry in same-sign
dilepton events with jets and missing
energy~\cite{Aad:2011vj,cmsnewer}, where the same-sign leptons arise
from the decay chains of the pair-produced gluino or squark particles'
cascade decays~\cite{sstheo}. Such a process is mediated by a
non-trivial color exchange in the $s$ or $t$ channels, which results
in large scattering angles of the energetic final state jets. This
signature, characterized by large $H_T=\sum_{i\in \text{jets}}
p_{T,i}$, is different from the typical WBF phenomenology. On the
other hand, since $\Br(H^\dpm\to W^\pm W^\pm)$ is large, we might
overcome the limitations of searches for light Higgs particles in
$H\to VV,~V=Z,W^\pm$, especially because the $H^\dpm W^\mp W^\mp$
coupling can be enhanced in comparison to $HW^+W^-$ due to the model's
triplet character. Furthermore, Ref.~\cite{cmsnewer}, which reports a
SUSY search employing the 7 TeV 4.98 fb$^{-1}$ data set, comprises
signal regions with relatively small $H_T\geq 80$~GeV (compensated
with a larger missing energy requirement) which can be exploited to
formulate constraints on the triplet model. This will be the focus of
Sec.~\ref{sec:cms}. Subsequently, in Sec.~\ref{sec:vbf7}, we
demonstrate that a slight modification of the search strategy of
Ref.~\cite{cmsnewer} is sufficient to obtain superior constraints on
the triplet model even for a pessimistic estimate of reducible
backgrounds and other uncertainties. We also discuss in how far these
estimates can be improved by including the 8~TeV data set. In
Sec.~\ref{sec:14tev} we discuss an analysis on the basis of a WBF
selection at $\sqrt{s} = 14$~TeV center-of-mass energy, which will
yield strong constraints on the triplet models' parameter space.

As we will argue, the results of these sections are not specific to a
particular triplet model and largely generalize to {\it any} model
with Higgs triplets. Since the tree-level custodial symmetry
preserving implementation of Higgs triplets exhibits a richer
phenomenology, we specifically analyze the impact of the described
searches in the context of the Georgi-Machacek (GM)
model~\cite{Georgi:1985nv} (which we quickly review in
Sec.~\ref{sec:mod} to make this work self-contained). In particular,
we input the direct search constraints for doubly charged scalars into
a global scan of the electroweak properties, also taking into account
EWPD. We give our summary in Sec.~\ref{sec:conc}.

\section{A consistent model of Higgs triplets}
\label{sec:mod}
The Georgi Machacek model~\cite{Georgi:1985nv} is a tree-level
custodial isospin-conserving implementation of Higgs triplets based on
scalar content
\begin{equation}
  \label{eq:higgsfields}
  \Phi=\left(\begin{matrix} \phi_2^{\ast} & \phi_1 \\ -\phi_1^{\ast} & \phi_2
    \end{matrix}\right), \quad 
  \Xi=\left(\begin{matrix} \chi_3^{\ast} & \xi_1 & \chi_1 \\
      -\chi_2^{\ast} & \xi_2 & \chi_2 \\
      \chi_1^{\ast} & -\xi_1^{\ast} & \chi_3 \\
    \end{matrix}\right)\,.
\end{equation}
$\Phi$ is a SM-like Higgs doublet necessary for introducing fermion
masses, and $\Xi$ combines the complex $(\chi_1,\chi_2,\chi_3)$ and
real $(\xi_1,\xi_2,-\xi_1^*)$ triplets such that an additional
$\SU(2)_R$ can act in the usual fashion ($\Xi\to U_L\Xi U^\dagger_R$
and $\Phi\to \tilde U_L\Phi \tilde U^\dagger_R$) leaving custodial
isospin unbroken after $\Phi$ and $\Xi$ obtain vacuum expectation
values (vevs) $\left\langle \Xi \right \rangle = v_\Xi \mathbbm{1}$,
$\left\langle \Phi \right \rangle = v_\Phi \mathbbm{1}$.

For the purpose of this paper we choose a Higgs sector Lagrangian
\begin{subequations}
  \label{eq:lag}
  \begin{multline}
    {\cal{L}}={1\over 2} \Tr \left[ D_{2,\mu} \Phi^\dagger D_2^\mu
      \Phi \right] + {1\over 2}\Tr \left[ {D}_{3,\mu} \Xi^\dagger
      {D}_3^\mu \Xi \right] - V(\Phi,\Xi) \\+ {\text{$\Phi$ Yukawa
        interactions}} \,,
  \end{multline}
  where we introduce the potential that triggers electroweak symmetry
  breaking
  \begin{multline}
    \label{eq:pot}
    V(\Phi,\Xi)={\mu_2^2\over 2} \tr \left( \Phi^c \Phi \right) +
    {\mu_3^2\over 2} \tr \left( \Xi^c \Xi \right) + \lambda_1
    \left[\tr \left( \Phi^c \Phi \right)\right]^2 \\ + \lambda_2 \tr
    \left( \Phi^c \Phi \right) \tr \left( \Xi^c \Xi \right) +
    \lambda_3 \tr \left( \Xi^c
      \Xi\, \Xi^c \Xi \right)\\
    + \lambda_4 \left[\tr \left( \Xi^c \Xi \right)\right]^2 -
    \lambda_5 \tr \left(\Phi^c t_2^a \Phi t_2^b \right) \tr \left(\Xi^c t_3^a
      \Xi t_3^b \right) \,.
  \end{multline}
\end{subequations}
This choice reflects the properties of the Higgs triplet model in a
simplified way \cite{Georgi:1985nv} and can be motivated from imposing
a ${\mathbb{Z}}_2$ symmetry \cite{chano}.
 
$D_2,D_3$ are the gauge-covariant derivatives in the $\SU(2)_L$
doublet and triplet representations. Hypercharge $\U(1)_Y$ is embedded
into $\SU(2)_R$ as in the SM, the ${\mathfrak{su}}(2)$ generators in
the triplet representation are
\begin{multline}
  t^1_3={1\over \sqrt{2}} \left( \begin{matrix}
      0 & 1 & 0 \\
      1 & 0 & 1 \\
      0 & 1 & 0 \end{matrix}\right)\,, \quad
  t^2_3={i\over \sqrt{2}} \left( \begin{matrix}
      0 & -1 & 0 \\
      1 &  0 & -1 \\
      0 &  1 & 0 \end{matrix}\right) \,,\\
  t^3_3=\left( \begin{matrix}
      1 & 0 & 0 \\
      0 & 0 & 0 \\
      0 & 0 & -1 \end{matrix}\right)\,.
\end{multline}

The masses of the electroweak bosons $m_W,m_Z$ after symmetry breaking
follow from the sum of the Higgs fields' vevs, constraining
\begin{equation}
  \label{eq:vev}
  (246~\hbox{GeV})^2= v_\Phi^2+8v_\Xi^2\,.
\end{equation}
Defining the mixing angles
\begin{equation}
  \label{eq:vevrot}
  \begin{split}
    \cos\theta_H =:&\;\,c_H={v_\Phi\over v_{\text{SM}}}\,, \\ 
    \sin\theta_H  =:&\;\,s_H={2\sqrt{2}v_\Xi\over v_{\text{SM}}}
  \end{split}
\end{equation}
turns out to be useful.  Since custodial isospin is preserved, in the
unitary gauge the Higgs masses group into two singlets, one triplet
and one quintet (the quintet includes our doubly charge scalar
$H_5^\dpm$, which we will indicate also without the subscript). Their
masses are
\begin{align}
  \begin{split}
    \label{eq:masses}
    m_{H_0}^2 &=  2 ( 2 \lambda_1 v_\Phi^2 + 2 ( \lambda_3 + 3 \lambda_4 ) v_\Xi^2 + m_{\Phi\Xi}^2 )\,, \\
    m_{H'_0}^2 &=  2 ( 2 \lambda_1 v_\Phi^2 + 2 ( \lambda_3 + 3 \lambda_4 ) v_\Xi^2 - m_{\Phi\Xi}^2 )\,, \\
    m_{H_3}^2&= {1\over 2} \lambda_5 (v_\Phi^2+8v_\Xi^2)\,,\\
    m_{H_5}^2&= {3\over 2} \lambda_5 v_\Phi^2 + 8 \lambda_3 v_\Xi^2\,,
  \end{split}
\end{align}
with short hand notation
\begin{multline}
  m_{\Phi\Xi}^2 = \Big{[} 4 \lambda_1^2 v_\Phi^4 - 8 \lambda_1
  (\lambda_3 +3 \lambda_4) v_\Phi^2 v_\Xi^2 \\
  + v_\Xi^2 \Big{(} 3 (2\lambda_2 - \lambda_5)^2 v_\Phi^2 + 4
  (\lambda_3 + 3 \lambda_4)^2 v_\Xi ^2 \Big{)} \Big{]}^{1/2}\,.
\end{multline}
To reach Eq.~\gl{eq:masses} we have diagonalized the singlet mixing by
an additional rotation
\begin{equation}
  \begin{split}
    H_0 &= \phantom{-} c_q H_\Phi + s_q H_\Xi\,, \\
    H'_0 &= -s_q H_\Phi + c_q H_\Xi \,,
  \end{split}
\end{equation}
with angle
\begin{multline}
  \label{eq:sq}
  \sin \angle(H_\Phi,H_0) =:s_q\\=\frac{\sqrt{3}}
  {\sqrt{3+\Big{[}\frac{2\lambda_1 v_\Phi^2 -2(\lambda_3
        +3\lambda_4)v_\Xi^2 + m_{\Phi\Xi}^2}{{(2\lambda_2 -
          \lambda_5)}v_\Phi v_\Xi}\Big{]}^2}}\,.
\end{multline}
Note that $m_{H'_0}< m_{H_0}$, and therefore $m_{H'_0}$ will be the
observed Higgs boson.

\begin{figure*}[!t]
  \subfigure[~Results of the CMS analysis of Ref~\cite{cmsnewer}. 
  The
  signal is computed from the
  $pp \to (H^\dpm_5  \to E_T^{\text{miss}}+\ell^+\ell^+ ) + jj$ process.
  The dashed vertical lines next to the bins give the background
    uncertainty in each search region (for details see text).]
  {\label{fig:cmschannels}
    \includegraphics[width=0.43\textwidth]{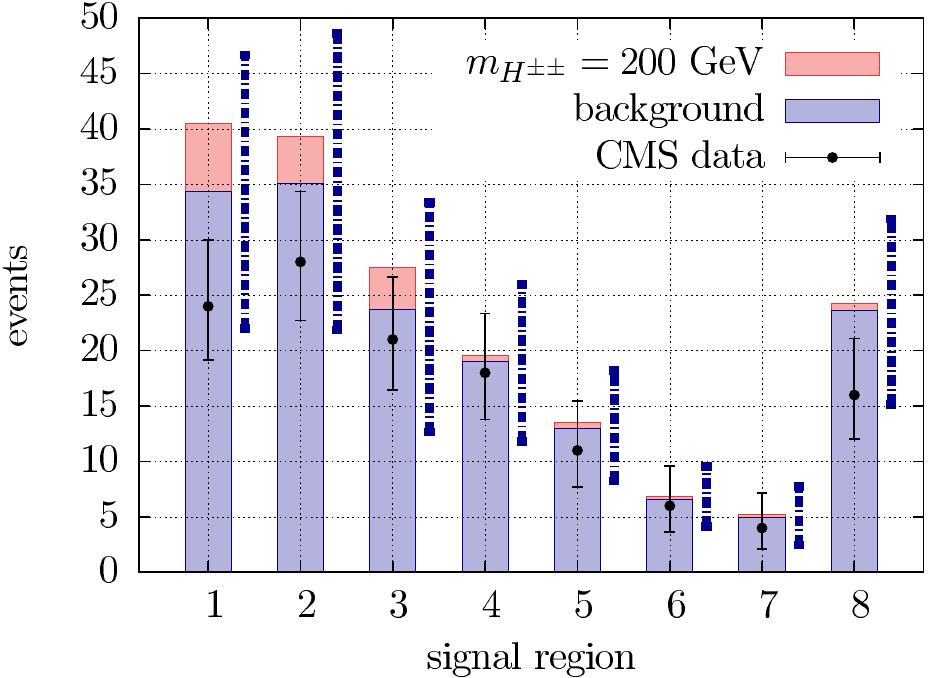}}
  \hspace{1cm} 
  \subfigure[~95\% CLS limits on the Georgi-Machacek model signal
  strength $\xi$ resulting from the 7 TeV selections of
  Ref.~\cite{cmsnewer}.]  {\label{fig:cmscls}
    \includegraphics[width=0.43\textwidth]{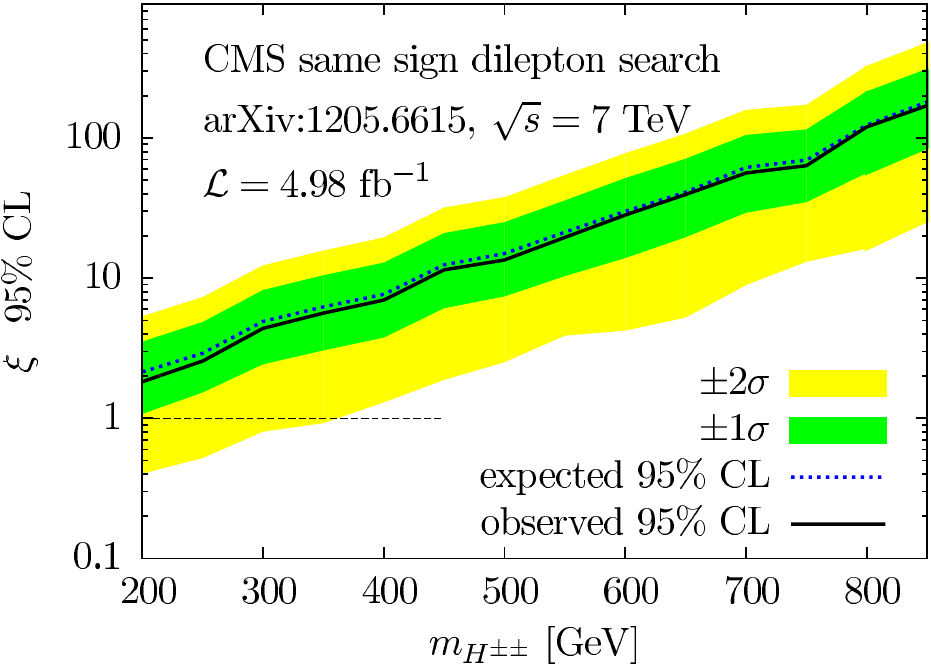}
  }
  \caption{\label{fig:cmsclsc} Estimated signal and background events,
    measured data and corresponding CLS limits for the 7 TeV CMS
    selection of Ref.~\cite{cmsnewer}.}
\end{figure*}

We straightforwardly compute the couplings of the uncharged states to
the SM fermions $f$ and gauge bosons $v$, normalized to the SM
expectation, as
\begin{equation}
  \begin{split}
    \label{eq:ctcv}
    c_{f,H_0} &= \frac{c_q}{c_H}\,, \nn \\
    c_{v,H_0} &= c_q c_H + \sqrt{8/3} \,s_q s_H\,, \nn\\
    c_{f,H'_0}&= -\frac{s_q}{c_H}\,, \nn\\
    c_{v,H'_0}&= -s_q c_H + \sqrt{8/3} \,c_q s_H\,.
  \end{split}
\end{equation}
The custodial triplet $(H_3^+,H_3^0,H_3^-)$ is gaugephobic and the
quintet fermiophobic with the additional assumption of a vanishing
leptonic Majorana operator. For the purpose of our analysis this does
not pose any phenomenological restriction. Since $\left\langle
  \Xi\right\rangle$ is the order parameter that measures the degree of
triplet symmetry breaking, a measurement of the $H^\dpm\to W^\pm
W^\pm$ directly reflects the phenomenology's triplet
character. Indeed, the vertex we are predominantly interested in is
given by
\begin{equation}
  \label{eq:dpvertex}
  H^\dpm W^\mp_\mu W^\mp_\nu: \quad \sqrt{2}i g m_W s_H g_{\mu\nu}\,,
\end{equation}
and, as we mentioned in Sec.~\ref{sec:intro}, the relevant final
states to study this vertex are therefore
``${E_T^{\text{miss}}}+\ell^\pm\ell^\pm$'' in association with at
least 2 jets. The $2$ jets signature will play the more important
role.

Note that Eq.~\gl{eq:dpvertex} implies that $H^\dpm$ can be enhanced
by up to a factor of two compared to the WBF production of a neutral
SM-like Higgs boson of the same mass. The enhanced couplings
Eqs.~\gl{eq:ctcv} and~\gl{eq:dpvertex} are a direct consequence of the
larger isospin of the triplet that feeds into the interactions via the
gauge kinetic terms.


At this stage it is important to comment on the relation of the
Georgi-Machacek model with ``ordinary'' triplet Higgs extension, {\it
  e.g.} when we just add a complex scalar field to the SM Higgs sector
with hypercharge $Y=2$ \cite{classic}. Such models introduce a
tree-level custodial isospin violation and consistency with EWPD
imposes a hierarchy of the vevs ($s_H\ll 1$). Since we are forced to
tune the model already at tree level the additional singly and doubly
charged states tend to decouple from the phenomenology apart from
loop-induced effects on branching ratios (see {\it e.g.}~\cite{andrew}
for reconciling the possibly observed excess in $H\to \gamma\gamma$ in
this fashion). The Georgi-Machacek model is {\it fundamentally}
different in this respect: due to the $SU(2)_R$ invariant extension of
the Higgs potential there are no tree-level constraints on $v_\Xi$. In
fact, only the generation of fermion masses requires the presence of
another doublet, and $2\sqrt{2} v_\Xi\gg v_\Phi$ does not lead to
tree-level inconsistencies in the gauge sector. At one loop, however,
this picture changes.  The presence of a triplet requires the explicit
breaking of $SU(2)_R$ invariance to tune the $\rho$ parameter to the
values consistent with EWPD~\cite{Gunion:1990dt,us} but still larger
values of $v_\Xi$ remain allowed in comparison to the simple complex
triplet extension, where recent upper bounds for the triplet vev read
as $v_{\text{triplet}} < 0.03 \times (246\hbox{ GeV})$~\cite{andrew}.

An analysis which measures $H^\dpm_5 \to W^\pm W^\pm$ is not specific
to the underlying model as Eq.~\gl{eq:dpvertex} simply follows from
the presence of a triplet Higgs in the particle spectrum that
contributes to electroweak symmetry breaking.  Since the Georgi
Machacek model accommodates larger values of $s_H$ with a rich
phenomenology we take this particular model as a benchmark for our
parameter fit in Sec.~\ref{sec:ewpd}. Our results generalize to any
triplet Higgs model implementation -- they provide constraints on this
branching ratio, which are model-independent statements as long as the
narrow width approximation can be justified.

\section{Re-interpreting SUSY searches}
\label{sec:cms}

We are now ready to compute an estimate of the performance of the CMS
analysis of Ref.~\cite{cmsnewer} when re-interpreted in the Higgs
triplet context. 

We focus on the light lepton flavor channel of Ref.~\cite{cmsnewer};
the additional $\tau$ lepton channels are subject to large fake
background uncertainties and do not provide statistical pull for our
scenario in the first place. The CMS analysis of Ref.~\cite{cmsnewer}
clusters anti-$k_T$ jets~\cite{antikt} with $R=0.5$ as implemented in
{\sc{FastJet}}~\cite{fastjet} and selects jets with $p_T> 40$ GeV in
$|\eta|<2.5$. Leptons are considered as isolated objects if the
hadronic energy deposit in within $\Delta R=[(\Delta \phi)^2+(\Delta
\eta)^2]^{1/2}=0.3$ is less than 15\% of the lepton candidate's
$p_T$. The thresholds are $p_{T,\mu}> 5~\gev$, and $p_{T,e}> 10~\gev$,
and there is a ``high $p_T$'' selection with $p_{T,\ell}> 10~\gev$
($\ell=e,\mu$) with the hardest lepton having $p_T> 20~\gev$. All
leptons need to fall within $|\eta|<2.4$. CMS requires at least two
jets and two leptons and vetos events with three leptons when one of
the leptons combines with one of the others to the $Z$ boson mass
within $\pm$15~GeV. CMS defines $H_T$ to be the scalar sum of all
jets' $p_T$ whose angular separation to the nearest lepton is $\Delta
R>0.4$.

We have generated CKKW-matched~\cite{ckkw} $t\bar t+W^\pm/Z$, $W^\pm
W^\pm jj$ and $W^\pm Zjj$ which constitute the dominant backgrounds
using {\sc{Sherpa}}~\cite{sherpa}.
The QCD corrections to theses processes are known to be
small~\cite{Campanario:2013qba,barbara,giulia,giuboz}. The signal
events are produced with {\sc{MadGraph}}/{\sc{MadEvent v5}}~\cite{mg5}
using a {\sc{FeynRules}}~\cite{Christensen:2008py} interface to our
model implementation described in Ref.~\cite{us}.\footnote{We note
  that the $2j+{E_T^{\text{miss}}}+\ell^+\ell^+$ signal also receives
  contributions from the vertices $H_0 W^\pm W^\pm$, $H'_0 W^\pm W^\pm
  $ and $H^0_5 W^\pm W^\pm$ which are present in diagrams containing
  an internal $t$-channel neutral Higgs boson connecting the $W^+$'s
  emitted from the two quark lines.  These diagrams are important to
  ensure unitarity in longitudinal weak boson scattering for high
  energy ($H_5^\dpm$ off-shell) scattering. We have checked that their
  numerical contribution is negligible in the $H_5^\dpm$ resonant
  region captured by Fig.~\ref{fig:graph}, and therefore we have not
  included them explicitly in this work.}
The signal events are subsequently showered and hadronized with
{\sc{Herwig++}}~\cite{herwig}. In the analysis we include gaussian
detector smearing of the jets and leptons on the basis of
Ref.~\cite{atlastdr}:
\begin{equation}
  \begin{split}
    \label{eq:resol}
    \text{jets}:\quad & {\Delta E\over E} = {5.2\over E} \oplus
    {0.16\over
      \sqrt{E}} \oplus 0.033\,,\\
    \text{leptons}:\quad & {\Delta {{E}} \over
      {E}} = {0.02}\,,\\
  \end{split}
\end{equation}
and we include the missing energy response from recent particle flow
fits of CMS~\cite{pflow} via the fitted
function~\cite{why}~\footnote{The missing energy response might vary
  from Ref.~\cite{pflow} when the analysis is performed by the
  experiments. This is clearly beyond the scope of this work, but we
  believe that our parametrization is well-justified for demonstration
  purposes.}
\begin{equation}
  \text{missing energy}:\quad  {\Delta {{E}}^{\text{miss}}_T \over
    {E}^{\text{miss}}_T} = {2.92 \over E^{\text{miss}}_T} - 0.07\,.
\end{equation}
The jet resolution parameters can be improved by particle flow too, we
however choose the more conservative parametrization to capture the
effect of an increased jet energy scale uncertainty in the forward
detector region, which especially impacts the WBF-like selection.

We use the background samples to generate an efficiency profile over
the 8 CMS search regions ({\it{cf.}}~Fig.~\ref{fig:cmschannels}) we
are focusing on
\begin{equation}
  \label{eq:search}
  \begin{split}
    \hbox{region 1:}\quad & {\text{high $p_T$}}, H_T>80~\gev,~E_T^{\text{miss}}>120~\gev,\\
    \hbox{region 2:} \quad & {\text{low $p_T$}}, H_T>200~\gev,~E_T^{\text{miss}}>120~\gev,\\
    \hbox{region 3:} \quad& {\text{high $p_T$}}, H_T>200~\gev,~E_T^{\text{miss}}>120~\gev,\\
    \hbox{region 4:} \quad& {\text{low $p_T$}}, H_T>450~\gev,~E_T^{\text{miss}}>50~\gev,\\
    \hbox{region 5:}\quad & {\text{high $p_T$}}, H_T>450~\gev,~E_T^{\text{miss}}>50~\gev,\\
    \hbox{region 6:} \quad& {\text{low $p_T$}}, H_T>450~\gev,~E_T^{\text{miss}}>120~\gev,\\
    \hbox{region 7:} \quad& {\text{high $p_T$}}, H_T>450~\gev,~E_T^{\text{miss}}>120~\gev,\\
    \hbox{region 8:}\quad & {\text{high $p_T$}}, H_T>450~\gev,~E_T^{\text{miss}}>0~\gev\,,
  \end{split}
\end{equation}
which we apply to our signal hypothesis.~\footnote{Since the CMS
  analysis does not tag on the number of jets, we have also considered
  production modes with same-sign dilepton and $E_T^{\text{miss}}$,
  but where more than 2 jets are produced. In a model with an extended
  Higgs sector, the cross section to produce such final states could
  potentially be very different from the SM rate. We have explicitly
  checked that production rates for $pp \to E_T^{\text{miss}}+
  \ell^\pm \ell^\pm + (> 2j)$ when extra states are included are
  negligible with respect to the main contribution to the signal,
  \emph{i.e.} $ pp\to W^\pm W^\pm jj$, with an $s$-channel exchanged
  $H_5^\dpm$, is the dominant process. To establish this, we have
  computed the impact of $pp\to Z\to H^\dpm H^\dmp\to W^\pm W^\pm
  jjjj$, $pp\ ( \to W^\pm ) \to H^\dpm H_{3,5}^\mp \to W^\pm W^\pm
  jjjj$, $pp\to H^\dpm \to W^\pm H_{3,5}^\pm \to W^\pm W^\pm jj$
  and $gg\to H_0\to H^\dpm H^\dmp\to W^\pm W^\pm jjjj$ ($gg\to H_3^0$
  would also be possible, but $H_3^0\to H^\dpm H^\dmp$ is forbidden)
  to the signal estimate, and found negligible contributions.  More
  precisely, the only process that could have a marginal impact is
  $gg\to H_0\to H^\dpm H^\dmp\to W^\pm W^\pm jjjj$, when $m_{H_0}>2
  m_{H_5^\dpm}$. While $H_0$ can be heavier than the quintet, the
  situation where it is heavy enough to have an open 2-body decay
  channel into a quintet pair is not very frequent. For example we
  have checked that this is the case by inspecting the points we
  considered in our previous study~\cite{us}. In the present work,
  only for the template scenarios with light quintets ($m_{H_5}<250$
  GeV) we have found that this is possible, and in such cases we have
  checked that the total contribution from this subprocess can enhance
  the signal by a factor $1.5$. This is not enough to change our
  estimates significantly.  We are therefore confident that the
  approximations we are using for the simulation of signal and
  backgrounds are robust. We however note that the contributions
  discussed in this footnote are model-dependent because they
  explicitly probe the larger particle content and the Higgs
  interactions due to the potential.}
To obtain CLS exclusion limits~\cite{Read:2002hq} we perform a log
likelihood hypothesis test as described in~\cite{junk}, where we
marginalize over the background uncertainty quoted in~\cite{cmsnewer}
(and indicated in Fig.~\ref{fig:cmschannels}).

The result is shown in Fig.~\ref{fig:cmscls}, where we plot the
observed and expected 95\% confidence level constraints on the signal
strength
\begin{equation}
  \xi=\frac{\sigma(H^\dpm jj)\times {\text{BR}}(H^\dpm \to W^\pm W^\pm
    \to {\text{leptons}})}{[\sigma(H^\dpm jj)\times {\text{BR}}(H^\dpm \to W^\pm W^\pm
    \to {\text{leptons}})]_{\text{ref}}}
\end{equation}
as function of the mass of $H^\dpm$. Since the total width is
dominated by $H_5^\dpm \to W^\pm W^\pm$, we have $\xi\simeq
s_H^2$. $\xi$ sets a limit in reference to a point that we choose with
values
\begin{equation}
  s_H=1/\sqrt{2}\,,~m_3=500~\gev
\end{equation}
for the Higgs mixing and triplet mass, {\it i.e.} a $hW^+W^-$-like
value of the $H^\dpm W^\pm W^\pm$ coupling.  These are also values
allowed by constraints from non-oblique corrections, in particular due
to $Z\to b\bar{b}$ measurements~\cite{Haber:1999zh}.  All other
parameters are chosen such that the 125 GeV Higgs state has a coupling
to weak gauge bosons that agrees with the SM within 5\%.
Given the large triplet Higgs vev, this is an optimistic scenario, but
we stress that it only serves to establish a baseline for the
measurement of $\xi$.

We see that the CMS analysis, which cuts on $H_T$, {\it i.e.}  central
jet activity instead of WBF-type topologies, only starts to probe
the model for $H^\dpm_5$ masses close to the $W^\pm$ threshold. The
discriminative power always predominantly comes from the search region
1, which is closest to a typical WBF selection among the eight search
channels of Eq.~\gl{eq:search}. As we will see in Sec.~\ref{sec:ewpd},
once other constraints such as electroweak precision measurements and
direct Higgs search constraints are included, the SUSY search does not
provide a strong constraint on the parameter space of the
Georgi-Machacek model.

As can be guessed from Fig.~\ref{fig:cmschannels}, excluding the
triplet via the CMS SUSY search is hampered by the large systematic
uncertainties. If we omit the systematic uncertainties and compute the
excluded signal strength only on the basis of statistical
uncertainties, the CMS analysis excludes $\xi=0.68$ for
$m_{H_5^\dpm}=200~\gev$. This enables a qualitative projection of the
situation when the 8 TeV sample is included. Due to the larger data
sample we can expect that the background uncertainty is reduced by a
larger available set of subsidiary background measurements at higher
statistics. CMS has an 8 TeV data sample of ${\cal{L}}\simeq
23~{\text{fb}}^{-1}$. With this sample and a systematic uncertainty
reduced by 50\%, CMS starts probing the triplet parameter space for
$H^\dpm_5$ masses up to $m_H^{\dpm}\simeq 250$~GeV.

\begin{figure}[!t]
  \includegraphics[width=0.43\textwidth]{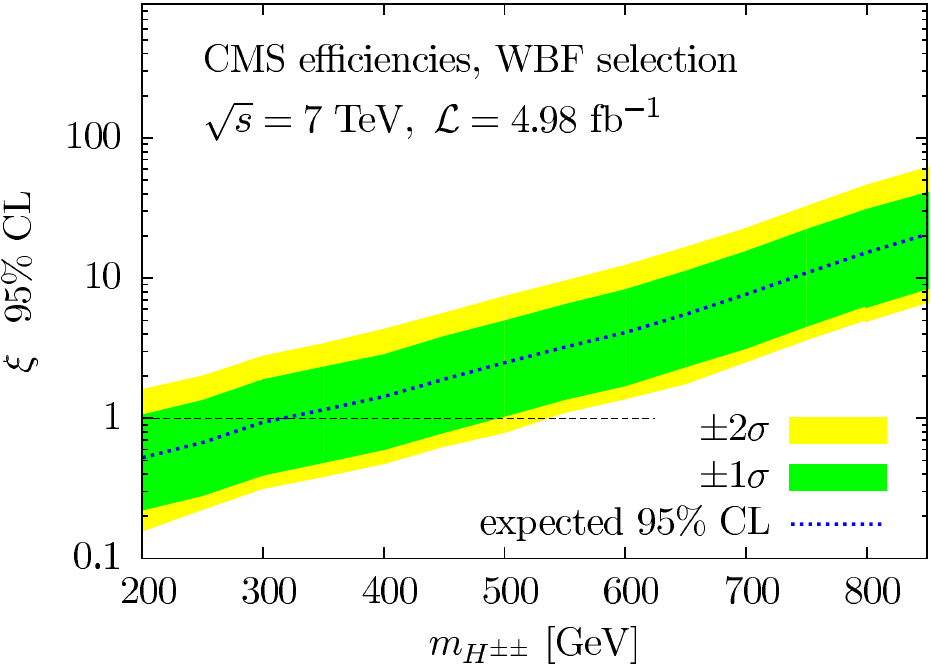}
  \caption{\label{fig:cmsclsvbf} Expected exclusion limits for a more
    WBF-like analysis based on Ref.~\cite{cmsnewer}. For details see
    text.}
\end{figure}

\begin{figure*}[!t]
  \subfigure[~Transverse cluster mass distribution.]{
    \label{fig:mt2c}\includegraphics[width=0.43\textwidth]{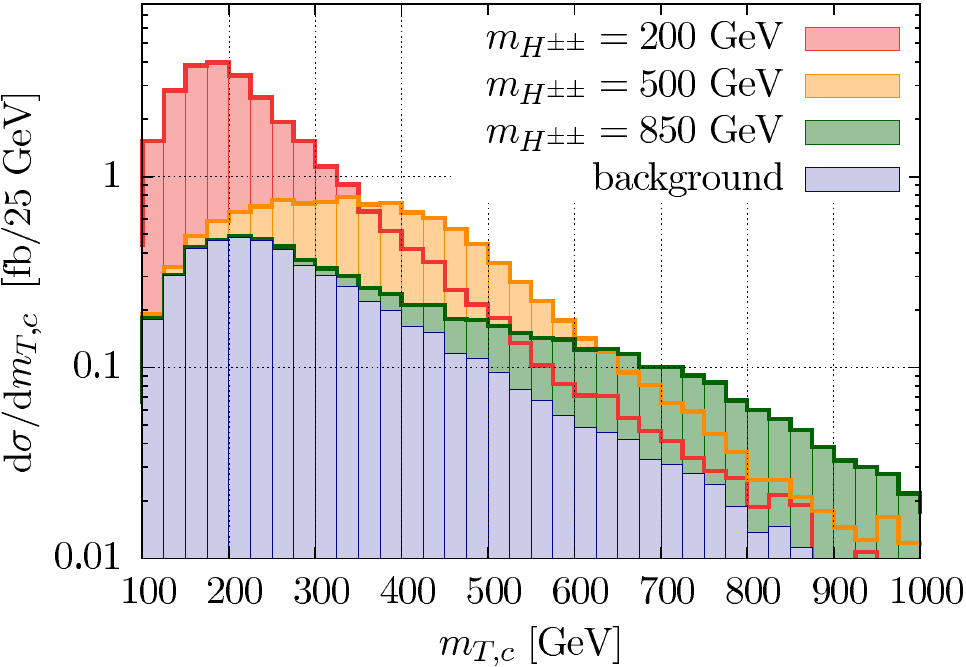}
  }\hspace{1cm} 
  \subfigure[~Normalized signal cluster mass distribution.]{
    \label{fig:mt2ci}\includegraphics[width=0.43\textwidth]{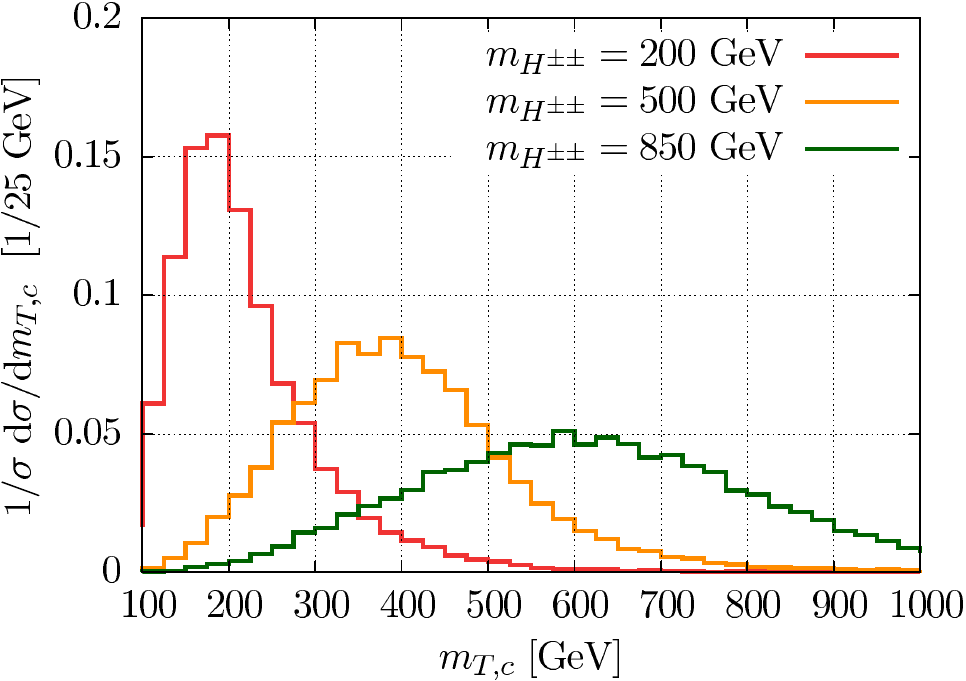}
  }
  \caption{\label{fig:invmassdis}Transverse cluster mass distribution
    for signal+background of $H_5^\dpm$ search as discussed in
    Sec.~\ref{sec:14tev}.  }
\end{figure*}

\subsection{Towards a more WBF-like selection}
\label{sec:vbf7}
We modify the above analysis towards a more signal-like selection. The
base cuts are identical, but this time we extend the jet clustering
over the full HCAL range $|\eta|<4.5$ and add standard WBF cuts via
\begin{equation}
  m_{j_1j_2}>500~\gev~\hbox{and}~|y_{j_1}-y_{j_2}|>4\,.
\end{equation}
This means that instead of exclusively clustering central jets, we
also allow more forward jets, so the systematic uncertainties might by
different compared to the CMS analysis we discussed above in
Sec.~\ref{sec:cms}. The fake background contribution, in particular,
can quantitatively only be assessed by the experiments themselves. To
get a qualitative estimate, we simulate $W+$heavy flavor
events\footnote{Following Ref.~\cite{cmsnewer} this is expected to be
  the dominant contribution of the fake background.} that we match
onto the CMS analysis region 1 and use a flat extrapolation to the
signal region described above. This yields approximately an estimate
of the background composition of again $\sim 60:40$ of
fake:irreducible. To calculate confidence levels we assume a
systematic uncertainty on the background of 75\% (which is a rather
conservative estimate in the light of the CMS search of the previous
section). As expected, the WBF selection reduces the background
without degrading the signal too much, therefore enhancing the signal
vs. background ratio. The expected exclusion limit on the basis of
these parameters is shown in Fig.~\ref{fig:cmsclsvbf}. We see that
already with the $4.98~{\text{fb}}^{-1}$ data set we can expect limits
on the model up to masses $m_{H^\dpm}\simeq 300~\gev$. If the
background uncertainty is reduced by 50\% the full 8 TeV data set
probes triplet models up to masses $m_{H^\dpm}\simeq 420~\gev$ for our
reference value $s_H=1/\sqrt{2}$.

\section{Prospective sensitivity and discovery thresholds at 14 TeV}
\label{sec:14tev}
Switching to higher center-of-mass energy changes the sensitivity to
the model dramatically. WBF-like cross sections increase by a factor
$\sim 5$ when doubling the available center-of-mass energy from 7 TeV
to 14 TeV~\cite{vbfnlo}. We can therefore introduce additional WBF
criteria like a central jet veto to further suppress the QCD
backgrounds, as well as lepton vetos to remove the $WZjj$ backgrounds.

Our event generation for the 14 TeV analysis follows the 7 TeV tool
chain. We use the anti-$k_T$ jets with $R=0.5$, and lower the $p_T$
thresholds to 20 GeV in $|\eta_j|<4.5$. We enlarge the requirement on
the tagging jets invariant mass to $m_{jj}>600~\gev$ and furthermore
require that the jets fall in opposite detector hemispheres
$y_{j_1}\cdot y_{j_2}\leq 0$. The leptons are required to be isolated
from the jets by a distance $\Delta R_{\ell j}=0.4$. This time we veto
events with a third lepton and a central jet which meets the above
requirement. No restrictions on $E_T^{\text{miss}}$ are imposed. The
result is a signal-dominated selection, which not only allows us to
highly constrain $s_H$ over a wide range of $H_5^\dpm$ masses but also
enables the approximate reconstruction of the $H_5^{\dpm}$ mass from a
Jacobian peak in the transverse cluster mass distribution,
\begin{widetext}
  \begin{equation}
    m_{T,c}^2=\left( \sqrt{(p_{\ell_1}+p_{\ell_2})^2+|\vec{p}_{T,\ell_1}+\vec{p}_{T,\ell_2}|^2} + 
      E_T^{\text{miss}}\right)^2 - 
    \Big{|}\vec{p}_{T,\ell_1}+\vec{p}_{T,\ell_2}+\vec{E}_{T}^{\text{miss}}\Big{|}^2
  \end{equation}
\end{widetext}
in case such a model is realized in nature. Due to detector resolution
effects, missing energy uncertainty and IS radiation, the mass
resolution of the Jacobian peak degrades significantly when
considering heavier $H^\dpm$ masses, as shown in
Fig.~\ref{fig:invmassdis}. A statistically significant measurement
will still be possible, the mass parameter determination, however,
will be poor.

The above event selection serves two purposes. Firstly, all
QCD-induced backgrounds (which are characterized by central jet
activity at moderate $m_{jj}$) are highly suppressed. We suppress the
backgrounds further by imposing lepton and central jet vetos. Note
that this also remove signal contributions which arise from other
processes other than WBF. As a result we directly constrain
$s_H$. After all cuts have been applied the irreducible background is
completely dominated by the electroweak SM $pp\to (W^\pm W^\pm\to
\ell^\pm\ell^\pm+E_T^{\text{miss}}) jj$ contribution at
${\cal{O}}(\alfa^6\alpha_s^0)$. This background is comparably small
and under good perturbative control~\cite{barbara}.

The fake background contribution can quantitatively only be assessed
by the experiments themselves. To get a qualitative estimate, we again
simulate $W+$heavy flavor events that we match onto the CMS analysis
region 1 and use a flat extrapolation to 14 TeV WBF selection
described above as already done for the 7 TeV WBF selection
criteria. This yields approximately an estimate of background
composition of 50:50 of fake:irreducible. We furthermore assume a
systematic shape uncertainty of the background of 35\% (flat), which
follows from 10\% and 25\% uncertainties on the irreducible and fake
background, respectively.
     
\begin{figure}[!t]
  \includegraphics[width=0.43\textwidth]{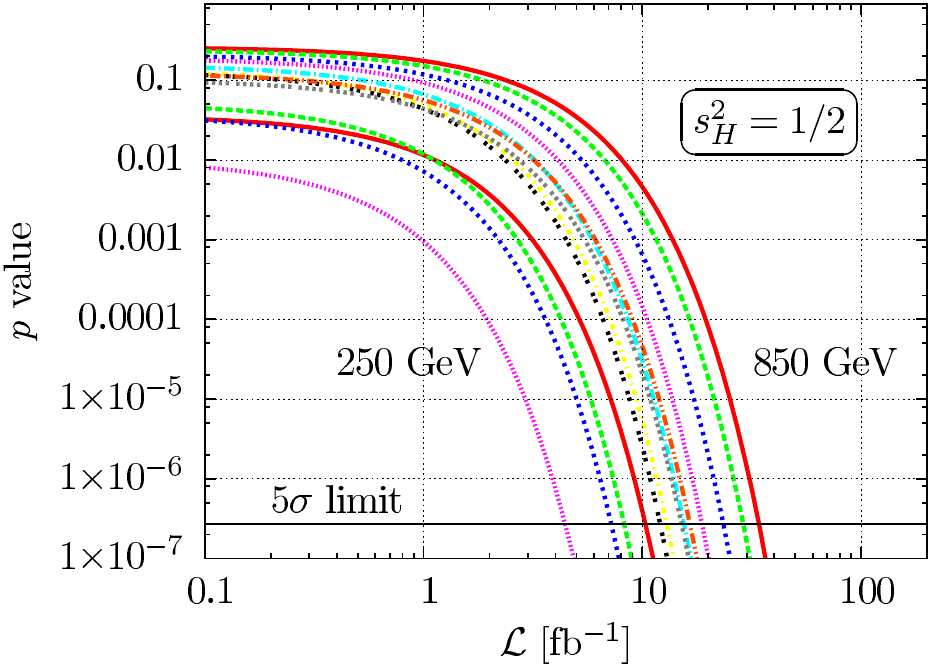}
  \caption{\label{fig:pval}Associated $p$ values for a search based on the single
    discriminant $m_{T,c}$ as a function of $m_{H^\dpm}$ and the
    integrated luminosity. The curves, moving from left to right,
    correspond to $H_5^\dpm$ masses between 250 GeV and 850 GeV in
    steps of 50 GeV.
  }
\end{figure}

In Fig.~\ref{fig:pval} we show the associated $p$ values for a search
based on the observable $m_{T,c}$ of Fig.~\ref{fig:invmassdis} for the
reference point $s_H=1/\sqrt{2}$. The signal cross section scales with
$ s_H^2 \sim s_H^{2\;\text{ref}}
\sqrt{{\cal{L}}/{\cal{L}}^\text{ref}}$. In principle this implies that
the LHC provides us enough sensitivity for discoveries down to
$s_H^2\sim 0.1$ for heavy masses for the considered $H_5^\dpm$ mass
range.

\begin{figure}[!t]
  \includegraphics[width=0.43\textwidth]{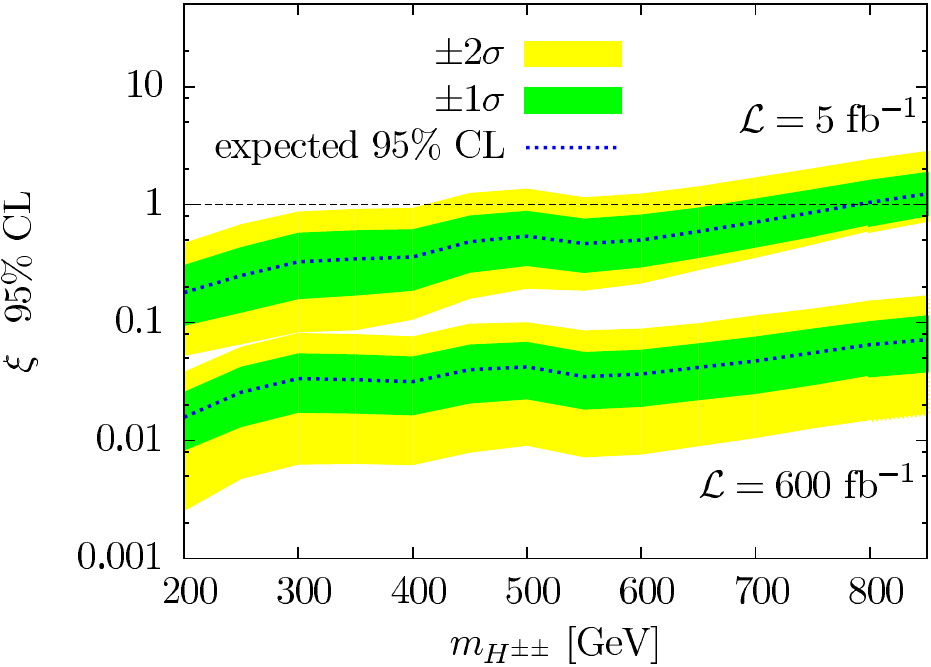}
  \caption{\label{fig:vbfcls14} Associated signal strength limits at
    95\% confidence level computed from a binned log likelihood
    hypothesis test on the basis of the single discriminant $m_{T,c}$,
    Fig.~\ref{fig:mt2c}, using the CL$_S$ method \cite{Read:2002hq}.
    We show results for two luminosity values for running at 14 TeV
    center-of-mass energy, 5~fb$^{-1}$ and 600~fb$^{-1}$.}
\end{figure}

In Fig.~\ref{fig:vbfcls14} we show the expected 95\% confidence level
constraints as a function of the $H^\dpm_5$ mass for luminosities
5~fb$^{-1}$ and 600~fb$^{-1}$. The expected constraint on the signal
strength $\xi$ can directly be interpreted as a limit on the $H_5^\dpm
W^\mp W^\mp$ coupling Eq.~\gl{eq:dpvertex}. As can be seen from this
figure, an analysis based on the WBF channel is a very sensitive
search, eventually yielding constraints $s_H^2\lesssim ~ 0.05$ over
the entire parameter range. On the one hand, since $s_H\ll 1$ is
required by the $W/Z$ mass ratio in a complex triplet extension the
expected constraint is not good enough to constrain the entire
parameter space. On the other hand, it is possible to constrain the
bulk of the parameter space in the context of the Georgi-Machacek
model, which typically allows larger values for $s_H$~\cite{us}.

\begin{figure*}[!t]
  \subfigure[~Higgs to diphoton branching ratio enhanced: $1.3\leq
  \xi_{H\to\gamma\gamma}\leq 2.3$]{
    \label{fig:modela}\includegraphics[width=0.43\textwidth]{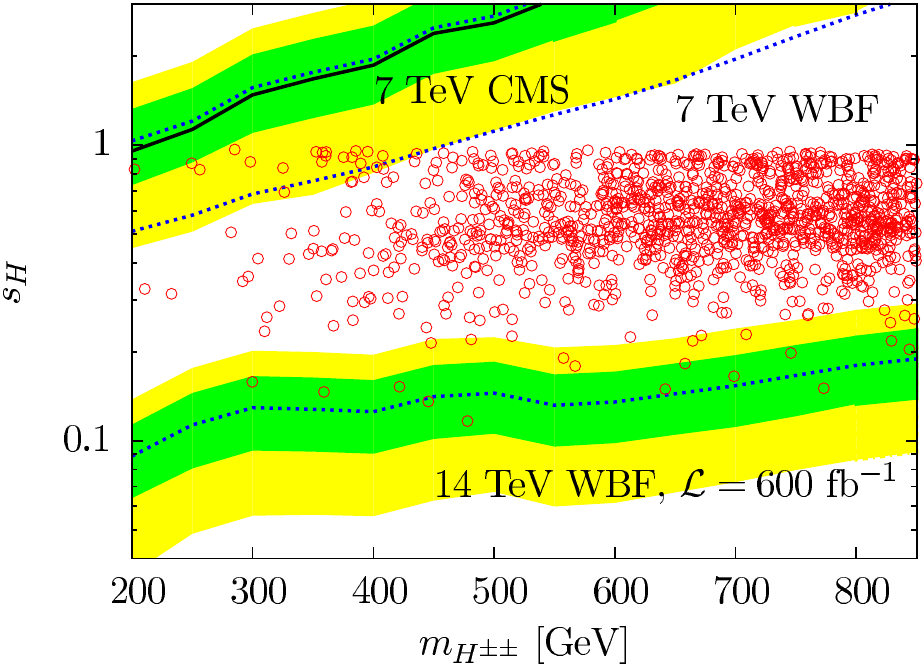}
  }\hspace{1cm} 
  \subfigure[~Higgs to diphoton branching ratio SM-like: $0.8\leq
  \xi_{H\to\gamma\gamma}\leq 1.2$]{
    \label{fig:modelb}\includegraphics[width=0.43\textwidth]{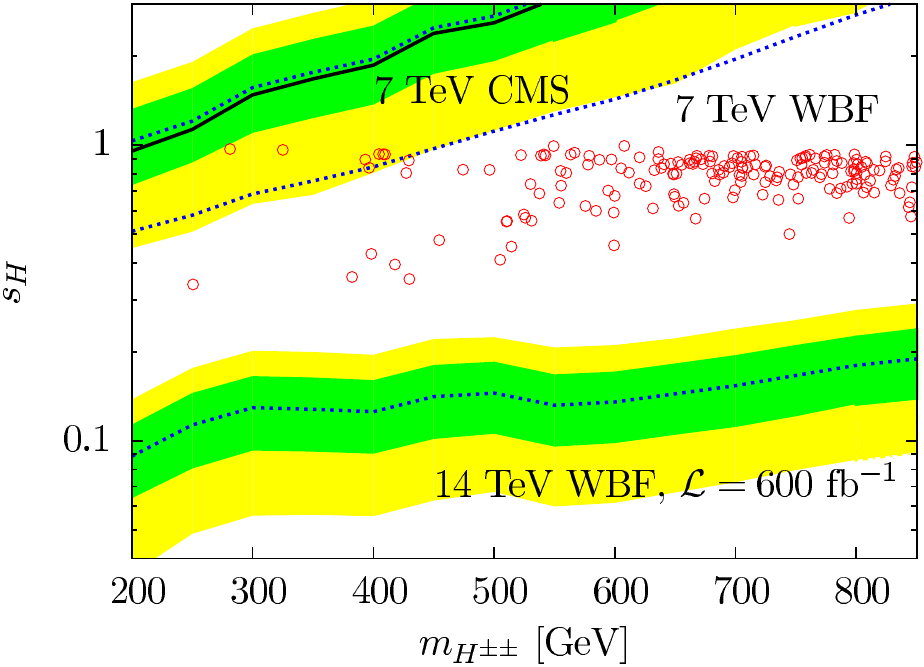}
  }
  \caption{Exclusion yield of the searches described in the previous
    sections when included to a model scan over the Georgi-Machacek
    model. The parameter points are consistent with electroweak
    precision measurements, current direct LHC and LEP constraints,
    and reproduce the signal strength of the measured Higgs boson in
    the observed weak boson decay channels $H\to W^+W^-,ZZ$. The
    dotted contours represent the expected exclusion, the green and
    yellow regions reflect the $\pm 1,\pm 2$ sigma uncertainty
    bands. The contours are, from top to bottom, the eight channel
    CMS SUSY search described in Sec.~\ref{sec:cms}, the adapted 7 TeV
    WBF search described in Sec.~\ref{sec:vbf7} (both
    $4.98~\text{fb}^{-1}$ luminosity) and the fully differential
    search at 14 TeV center of mass energy of Sec.~\ref{sec:14tev}
    ($600~\text{fb}^{-1}$ luminosity).}
\end{figure*}

\section{Combining direct $H_5^\dpm$ searches with other constraints}
\label{sec:ewpd}

In this section we want to compare the exclusion potentials due to
searches for a doubly charged scalar obtained in the previous sections
with representative points for the parameter space still allowed for
the GM model. In particular, in a previous study~\cite{us}, we have
shown that this space is large enough to accommodate both the case
where the 125 GeV Higgs boson has an enhanced $\gamma\gamma$ decay
rate with respect to the SM value and the case where the couplings for
the Higgs boson candidate are SM-like.

It is therefore natural to study if the (future, possible)
non-observation of excesses in searches for doubly charged states has
the potential to completely rule out these two scenarios, and hence
the GM extension of the Higgs sector.

Before showing the results, we summarize the information included in
the two sets of points we will use in the following. We list here only
the aspects which are relevant for the present work, and we refer the
reader to Ref.~\cite{us} for a detailed explanation of how these
results were obtained:

\begin{cedescription}
\item[Direct ATLAS, CMS] The points that we consider correspond to
  scenarios where the $H'_0$ scalar is the observed Higgs
  boson. Therefore we restrict to the case where the other singlet
  $H_0$ is heavier, and we require that neither $H_0$ nor $H_3^0$
  violate the LHC exclusion limits on scalar production. This case has
  been discussed in Ref.~\cite{us} in detail.
\item[Consistency with 125~GeV signal] We require that the tree-level
  couplings of $H'_0$ with fermions and gauge bosons, and the
  loop-induced coupling with gluons, are such that $H'_0$ reproduce
  the observed total signal strength as well as the individual signal
  strengths for $WW$ ($\xi_{H\to WW}$) and $\gamma\gamma$
  ($\xi_{H\to\gamma\gamma}$) decays. In particular, at this level we
  distinguish among a scenario where we have room to reproduce an
  excess in the photonic branching ratio and another where signal
  strengths agree with the SM values within 20\%. For further details
  on the scan we refer the reader to Ref.~\cite{us}.
\item[Oblique corrections] In our previous study we have also taken
  into account constraints from electroweak precision measurements. In
  particular we studied both cases where the $T$ parameter is used
  or not, since at one-loop the radiative corrections are not
  unambiguously defined.  In this work we have decided not to consider
  this subtle but important issue, which we instead discussed at
  length in Ref.~\cite{us}: therefore we used the sets of points
  labelled in our previous paper as ``S. param included'', {\it i.e.}
  the results obtained here are independent of any $T$ parameter
  constraint or fine tuning \cite{Gunion:1990dt}.
\item[Non-oblique corrections ($Zb\bar b$)] In our previous work we
  have not explicitly included constraints due to the fermionic
  coupling of the custodial-triplet charged states $H_3^\pm$. The
  presence of these states might change significantly several
  observables involving $b$-quarks, because of possibly large values
  for the $H_3^+ t b$ coupling. One of the more important observables
  to look at is $R_b$, defined as $\Gamma(Z\to
  b\bar{b})/\Gamma(Z\to\mbox{hadrons})$. Changes in the SM value
  prediction of $R_b$ induced by the GM model have been computed in
  Ref.~\cite{Haber:1999zh}.  We have reproduced these results, and
  checked that a large portion of the points we will use in the
  following, that were considered still allowed in our previous paper,
  survive also the bounds from $Z\to b\bar{b}$.~\footnote{For the sake
    of completeness, we would like to point out that recent results in
    the computation of 2-loop corrections for the SM $Zb\bar{b}$
    coupling lead to sizeable effects which have not been taken into
    account in previous literature~\cite{Freitas:2012sy}. Including
    these effects goes however beyond the purpose of this study,
    although it could be potentially relevant for constraints only due
    to non-oblique corrections. We will however show that searches for
    WBF-produced doubly charged states are very powerful as exclusion
    tests for these models, and therefore our main results will hold,
    regardless of the relative size of these loop effects.}
\end{cedescription}

As the above discussion shows, in our previous study we have taken
into account essentially all the available constraints from direct and
indirect searches. In particular, for this paper we also checked that
the conclusions we reached in Ref.~\cite{us} remain essentially
unchanged also when non-oblique corrections (in the $Z\to b\bar{b}$
case) are included.

In Figs.~\ref{fig:modela} and~\ref{fig:modelb} we show the exclusion
potential of the search strategies discussed in Sec.~\ref{sec:cms}
and~\ref{sec:14tev}, together with the surviving points for the two
scenarios we just described. The standard color coding is used for the
exclusion plots, which here are shown as a function of $s_H$ and
$m_{H^\dpm}$. From these plots we conclude that the searches at 7 TeV,
if extended with WBF-like selection cuts, start to be able to probe,
and hence exclude, some of the surviving scenarios.  The more relevant
result, however, is that WBF searches on the 14 TeV data will have the
potential to completely rule out all the points that survive all other
constraints. This search has therefore the potential to become
a decisive obstacle that models with Higgs triplets and large
triplet-doublet mixing have to pass in order not to be excluded. As
such, it would be very important for LHC experimental collaborations
to look into these final states. In particular an analysis based on
the same-sign lepton WBF channel serves to also constrain the
parameter region which is allowed in other recent analyses such as
Ref.~\cite{Belanger:2013xza}.

\section{Summary}
\label{sec:conc}
Higgs triplets as implemented in the Georgi-Machacek Model provide a
viable extension of the SM Higgs sector which can be efficiently
probed at the LHC. We have demonstrated that while current analyses of
same-sign lepton final states do not provide a strong enough
constraint on the presence of doubly charged scalar bosons decaying to
same-sign $W$'s on the basis of SUSY searches, the enlarged
statistical sample of the 8 TeV 2012 run should start constraining
this model via the non-adapted SUSY search strategy. Furthermore we
have shown that a simple modification of these SUSY searches allows us
to constrain the model already with 7 TeV data even for a conservative
background estimate. The model can be ultimately verified or ruled out
at the LHC with 14~TeV in a clean WBF selection.

Our results are quite general and at the same time realistic as far as
models with triplets are considered. In particular, studying
$H^\dpm\to W^\pm W^\pm$ rather than the more commonly considered case
$H^\dpm\to\ell^\pm\ell^\pm$ seems to be more natural, because of the
dominance of the former decay over the latter for the bulk of the
parameter space independent of the considered triplet
scenario. Moreover, the analysis strategies we studied in this work
are quite standard, but at the same time can lead to conclusive
results for a complete exclusion of Higgs sectors with triplets.  We
therefore think that it would be very important for the LHC
experimental collaborations to consider these searches in addition to
the already considered and simpler case $H^\dpm\to\ell^\pm\ell^\pm$.

\acknowledgments 
CE acknowledges funding by the Durham International
Junior Research Fellowship scheme. We also acknowledge Chris Hays for
interesting discussions, and for comments on the draft.



\begin{thebibliography}{99}

\bibitem{:2012gk}
  The ATLAS collaboration,
  Phys.\ Lett.\ B {\bf 716} (2012) 1.

\bibitem{:2012gu}
  The CMS collaboration,
  Phys.\ Lett.\ B {\bf 716} (2012) 30.

\bibitem{newboundsa}
  The ATLAS collaboration, ATLAS-CONF-2012-170.

\bibitem{newboundsb}
  The CMS collaboration, CMS-PAS-HIG-12-045.

\bibitem{orig} 
  F.~Englert and R.~Brout,
  Phys.\ Rev.\ Lett.\  {\bf 13} (1964) 321.
  P.~W.~Higgs,
  Phys.\ Lett.\  {\bf 12} (1964) 132 and
  Phys.\ Rev.\ Lett.\  {\bf 13} (1964) 508.
  G.~S.~Guralnik, C.~R.~Hagen and T.~W.~B.~Kibble,
  Phys.\ Rev.\ Lett.\  {\bf 13} (1964) 585.

\bibitem{lit}
  A.~Arhrib, R.~Benbrik, M.~Chabab, G.~Moultaka and L.~Rahili,
  JHEP {\bf 1204}, 136 (2012).
  S.~Chang, C.~A.~Newby, N.~Raj and C.~Wanotayaroj,
  Phys.\ Rev.\ D {\bf 86} (2012) 095015.
  L.~Wang and X.~-F.~Han,
  arXiv:1303.4490 [hep-ph].

\bibitem{andrew}
  A.~G.~Akeroyd and S.~Moretti,
  Phys.\ Rev.\ D {\bf 86} (2012) 035015.
  L.~Wang and X.~-F.~Han,
  Phys.\ Rev.\ D {\bf 87} (2013) 015015.

\bibitem{us}
  C.~Englert, E.~Re and M.~Spannowsky,
  Phys.\ Rev.\ D {\bf 87}, 095014 (2013).

\bibitem{Killick:2013mya}
  R.~Killick, K.~Kumar and H.~E.~Logan,
  arXiv:1305.7236 [hep-ph].

\bibitem{cmsaa}
  The CMS collaboration, CMS-PAS-HIG-13-001.

\bibitem{Georgi:1985nv}
  H.~Georgi and M.~Machacek,
  Nucl.\ Phys.\  B {\bf 262} (1985) 463.
  
\bibitem{chano}
  M.~S.~Chanowitz and M.~Golden,
  Phys.\ Lett.\ B {\bf 165} (1985) 105.

\bibitem{carmi}
  D.~Carmi, A.~Falkowski, E.~Kuflik, T.~Volansky and
  J.~Zupan,
  JHEP {\bf 1210}, 196 (2012).

\bibitem{Belanger:2013xza}
  G.~Belanger, B.~Dumont, U.~Ellwanger, J.~F.~Gunion and S.~Kraml,
  arXiv:1306.2941 [hep-ph].

\bibitem{classic}
  W.~Konetschny and W.~Kummer,
  Phys.\ Lett.\ B {\bf 70} (1977) 433.
  J.~Schechter and J.~W.~F.~Valle,
  Phys.\ Rev.\ D {\bf 22} (1980) 2227.
  T.~P.~Cheng and L.~-F.~Li,
  Phys.\ Rev.\ D {\bf 22} (1980) 2860.
  A.~Hektor, M.~Kadastik, M.~Muntel, M.~Raidal and L.~Rebane,
  Nucl.\ Phys.\ B {\bf 787} (2007) 198.

\bibitem{searchhpp}
  The CMS collaboration,
  arXiv:1207.2666 [hep-ex].
  The ATLAS collaboration,
  arXiv:1210.5070 [hep-ex].

\bibitem{Gunion:1989ci} 
  J.~F.~Gunion, R.~Vega and J.~Wudka,
  Phys.\ Rev.\ D {\bf 42}, 1673 (1990).

\bibitem{Kanemura:2013vxa}
  S.~Kanemura, K.~Yagyu and H.~Yokoya,
  arXiv:1305.2383 [hep-ph].

\bibitem{dieter}
  D.~L.~Rainwater and D.~Zeppenfeld,
  Phys.\ Rev.\ D {\bf 60} (1999) 113004
  [Erratum-ibid.\ D {\bf 61} (2000) 099901],
  N.~Kauer, T.~Plehn, D.~L.~Rainwater and D.~Zeppenfeld,
  Phys.\ Lett.\ B {\bf 503} (2001) 113,
  T.~Figy, C.~Oleari and D.~Zeppenfeld,
  Phys.\ Rev.\ D {\bf 68} (2003) 073005,
  C.~Oleari and D.~Zeppenfeld,
  Phys.\ Rev.\ D {\bf 69} (2004) 093004,
  B.~Jager, C.~Oleari and D.~Zeppenfeld,
  Phys.\ Rev.\ D {\bf 73} (2006) 113006,
  B.~Jager, C.~Oleari and D.~Zeppenfeld,
  JHEP {\bf 0607} (2006) 015,
  M.~Ciccolini, A.~Denner and S.~Dittmaier,
  Phys.\ Rev.\ Lett.\  {\bf 99} (2007) 161803,
  M.~Ciccolini, A.~Denner and S.~Dittmaier,
  Phys.\ Rev.\ D {\bf 77} (2008) 013002,

\bibitem{barbara}
  B.~Jager, C.~Oleari and D.~Zeppenfeld,
  Phys.\ Rev.\ D {\bf 80} (2009) 034022.
  B.~Jager and G.~Zanderighi,
  JHEP {\bf 1111} (2011) 055.

\bibitem{giulia}
  T.~Melia, K.~Melnikov, R.~Rontsch and G.~Zanderighi,
  JHEP {\bf 1012} (2010) 053.

\bibitem{Chiang}
 C.~-W.~Chiang, T.~Nomura and K.~Tsumura,
  Phys.\ Rev.\ D {\bf 85} (2012) 095023.
  C.~-W.~Chiang and K.~Yagyu,
  JHEP {\bf 1301} (2013) 026.

\bibitem{vbftriplet}
  S.~Godfrey and K.~Moats,
  Phys.\ Rev.\  D {\bf 81}, 075026 (2010).
  K.~Cheung and D.~K.~Ghosh,
  JHEP {\bf 0211} (2002) 048.


\bibitem{Aad:2011vj}
 The CMS collaboration,
  Phys.\ Rev.\ Lett.\  {\bf 109}, 071803 (2012) and
  CMS-PAS-SUS-12-017.
  The ATLAS collaboration,
  JHEP {\bf 1110} (2011) 107.

\bibitem{cmsnewer}
S.~Chatrchyan {\it et al.}  [CMS Collaboration],
  Phys.\ Rev.\ Lett.\  {\bf 109} (2012) 071803

\bibitem{sstheo}
 R.~M.~Barnett, J.~F.~Gunion and H.~E.~Haber,
  Phys.\ Lett.\ B {\bf 315} (1993) 349.

\bibitem{Gunion:1990dt} 
  J.~F.~Gunion, R.~Vega and J.~Wudka,
  Phys.\ Rev.\ D {\bf 43}, 2322 (1991).

\bibitem{antikt}
  M.~Cacciari, G.~P.~Salam and G.~Soyez,
  JHEP {\bf 0804} (2008) 063.

\bibitem{fastjet}
  M.~Cacciari, G.~P.~Salam and G.~Soyez,
  Eur.\ Phys.\ J.\ C {\bf 72} (2012) 1896.

\bibitem{ckkw}
  S.~Catani, F.~Krauss, R.~Kuhn and B.~R.~Webber,
  JHEP {\bf 0111} (2001) 063.

\bibitem{sherpa}
  T.~Gleisberg, S.~Hoeche, F.~Krauss, M.~Schonherr, S.~Schumann, F.~Siegert and J.~Winter,
  JHEP {\bf 0902}, 007 (2009).

\bibitem{giuboz}
  G.~Bozzi, B.~Jager, C.~Oleari and D.~Zeppenfeld,
  Phys.\ Rev.\ D {\bf 75} (2007) 073004.

\bibitem{Campanario:2013qba}
  F.~Campanario, M.~Kerner, L.~D.~Ninh and D.~Zeppenfeld,
  arXiv:1305.1623 [hep-ph].

\bibitem{mg5}
 J.~Alwall, M.~Herquet, F.~Maltoni, O.~Mattelaer and T.~Stelzer,
  JHEP {\bf 1106} (2011) 128.

\bibitem{Christensen:2008py}
  N.~D.~Christensen and C.~Duhr,
  Comput.\ Phys.\ Commun.\  {\bf 180} (2009) 1614.

\bibitem{herwig}
M.~Bahr, S.~Gieseke, M.~A.~Gigg, D.~Grellscheid, K.~Hamilton, O.~Latunde-Dada, S.~Platzer and P.~Richardson {\it et al.},
  Eur.\ Phys.\ J.\ C {\bf 58} (2008) 639.

\bibitem{atlastdr}
  G.~Aad {\it et al.}  [ATLAS Collaboration],
  JINST {\bf 3} (2008) S08003;
 G.~L.~Bayatian {\it et al.}  [CMS Collaboration],
  J.\ Phys.\ G {\bf 34} (2007) 995.

\bibitem{pflow}
  The CMS collaboration, CMS PAS PFT-09/001.

\bibitem{why}
  C.~Englert, M.~Spannowsky and C.~Wymant,
  Phys.\ Lett.\ B {\bf 718} (2012) 538.

\bibitem{Read:2002hq}
  A.~L.~Read,
  CERN-OPEN-2000-205.
  A.~L.~Read,
  J.\ Phys.\ G {\bf G28 } (2002)  2693-2704.
  G.~Cowan, K.~Cranmer, E.~Gross and O.~Vitells,
  Eur.\ Phys.\ J.\ C {\bf 71}, 1554 (2011).

\bibitem{junk}
  T.~Junk,
  Nucl.\ Instrum.\ Meth.\  A {\bf 434} (1999) 435.
  T.~Junk,
  CDF Note 8128 [cdf/doc/statistics/public/8128].
  T.~Junk,
  CDF Note 7904 [cdf/doc/statistics/public/7904].
  H. Hu and J. Nielsen, 
  in 1st Workshop on Confidence Limits’, 
  CERN 2000-005 (2000).

\bibitem{Haber:1999zh}
  H.~E.~Haber and H.~E.~Logan,
  Phys.\ Rev.\ D {\bf 62} (2000) 015011.

\bibitem{vbfnlo}
  K.~Arnold, M.~Bahr, G.~Bozzi, F.~Campanario, C.~Englert, T.~Figy, N.~Greiner and C.~Hackstein {\it et al.},
  Comput.\ Phys.\ Commun.\  {\bf 180} (2009) 1661.

\bibitem{Freitas:2012sy} 
  A.~Freitas and Y.~-C.~Huang,
  JHEP {\bf 1208}, 050 (2012)
  [Erratum-ibid.\  {\bf 1305}, 074 (2013)]

\end{thebibliography}
\end{document}